\documentclass[%
reprint,
superscriptaddress,
amsmath,amssymb,
aps,
prl,
]{revtex4-1}
\usepackage{graphicx}
\usepackage{dcolumn}
\usepackage{bm}
\usepackage{amsmath}
\usepackage{hyperref}
\hypersetup{hypertex=true,colorlinks=true,linkcolor=blue,anchorcolor=blue,citecolor=blue,urlcolor=blue}
\usepackage{lineno}
\usepackage{enumerate}
\usepackage{color}
\usepackage{upgreek}
\usepackage{threeparttable}
\usepackage{multirow}
\begin{document}
\preprint{APS/123-QED}

\title{Large-Angle Collisions in Burning Plasmas of Inertial Confinement Fusions}

\author{Y. H. Xue}
\affiliation{Key Laboratory for Laser Plasmas and Department of Physics and Astronomy, Collaborative Innovation Center of IFSA (CICIFSA), Shanghai Jiao Tong University, Shanghai 200240, People’s Republic of China}

\author{D. Wu}
\email{dwu.phys@sjtu.edu.cn}
\affiliation{Key Laboratory for Laser Plasmas and Department of Physics and Astronomy, Collaborative Innovation Center of IFSA (CICIFSA), Shanghai Jiao Tong University, Shanghai 200240, People’s Republic of China}

\author{J. Zhang}
\email{jzhang1@sjtu.edu.cn; jzhang@iphy.ac.cn}
\affiliation{Key Laboratory for Laser Plasmas and Department of Physics and Astronomy, Collaborative Innovation Center of IFSA (CICIFSA), Shanghai Jiao Tong University, Shanghai 200240, People’s Republic of China}
\affiliation{Institute of Physics, Chinese Academy of Sciences, Beijing 100190, People’s Republic of China}%

\date{\today}

\begin{abstract}
A recent neutron analysis of experiments conducted at the National Ignition Facility (NIF) has revealed deviations from the Maxwellian distributions in the ion relative kinetic energy of burning plasmas, with the surprising emergence of supra-thermal deuterium and tritium (DT) ions that fall outside the predictions of macroscopic statistical hydrodynamic models. Our hybrid-particle-in-cell simulations, incorporating the newly-developed model of large-angle collisions, suggest this could be attributed to the increased significance of large-angle collisions among DT ions and \(\alpha\)-particles in the burning plasma. Extensive investigations into the implications of large-angle collisions in the burning plasma have yield several key findings, including an ignition moment promotion by \(\sim 10\, {\rm ps}\), the presence of supra-thermal ions below an energy threshold, and a hotspot expansion rate about six times faster than expected. Furthermore, we have established the congruency between the NIF neutron spectral moment analysis and our simulations. Our researches on large-angle collisions in burning plasmas offer new insights for experiment interpretation and update our understanding for new designs of inertial confinement fusions.
\end{abstract}

\maketitle
\textit{Introduction}\label{sec:first}
Over the last few years, breakthroughs have been achieved in inertial confinement fusion (ICF). Remarkably, ignition was achieved on August 8th, 2021 \cite{abu2022lawson}, and on December 4th, 2022, a target gain of \(G>1\) was realized \cite{abu2024achievement}. Actually, these above achievements are rooted in a milestone achieved in February 2021: the attainment of the burning plasma state, where fusion self-heating surpassed the mechanical work injected into the implosions \cite{zylstra2022burning}. In this regime, novel physics emerges, especially kinetic effects. For example, Hartouni \textit{et al} recently reported the observation of ion relative kinetic energy distributions in burning plasmas that deviate from Maxwellian distributions, along with unexpected supra-thermal deuterium and tritium (DT) ions \cite{hartouni2022evidence}.

The deposition of supra-thermal \(\alpha\)-particles, with mean free paths exceeding characteristic length scales of typical ICF targets, invalidates the underlying hypothesis of fluid descriptions \cite{murphy2014effect}. In such burning plasmas, direct Coulomb collisions between high-energy DT ions and fusion \(\alpha\)-particles, especially large-angle collisions, are crucial in the \(\alpha\)-particles deposition. Large-angle collisions, which exchange substantial energy through a single collision, could generate a group of supra-thermal ions, providing insights into departures from the Maxwellian hydrodynamic behaviour in Hartouni's experiments. Conversely, considering only small-angle collisions leads to a Maxwellian distribution due to multiple ion collisions.

Modelling Coulomb collisions in burning plasmas poses a great challenge for ICF studies \cite{Andrade1984Nuclear, Fisher1994Fast, Gorini1995Alpha, Ballabio1997alpha, Kallne2000Observation, Molvig2012Knudsen, Albright2013Revised, Frenje2013Diagnosing}.
Frameworks favouring small-angle collisions have been developed and refined with theories of cumulative small-angle collisions \cite{takizuka1977binary, nanbu1997theory}. 
Large-angle collisions occur when the impact parameter satisfies \(b<b_c\), a pivotal threshold that distinguishes them from small-angle collisions by delineating the boundary between ``remote'' small-angle and ``close'' large-angle collisions. The self-consistent determination of \(b_c\) remains an open issue and is currently approximated based on a static model with screened Coulomb potentials \cite{turrell2014effects}. However, neutron spectral moments are directly linked to the reactant kinematics \cite{crilly2022constraints}, where the relative velocity of ions plays a significant role in influencing the shift in the mean kinetic energy of neutrons. Therefore, current model necessitates refinement to account for the relative motion of ions. 

In this Letter, we introduce a new model that incorporates large-angle collisions to comprehensively capture ion kinetics. This model integrates the screened potentials of the background with the relative motion of ions during binary collisions, which has been benchmarked by first-principle molecular dynamics (MD) simulations with an unprecedented number of particles, specifically at the scale of millions. Leveraging this model, we present findings obtained from our advanced hybrid-PIC code, which simulates the behavior of burning plasma under experimental conditions. Our analysis reveals the significant impact of large-angle collisions on various crucial aspects, including the promotion of the ignition moment by \(\sim 10\, {\rm ps}\), the supra-thermal distribution beneath a specific energy threshold before the ignition moment, and the deposition pattern of \(\alpha\)-particles near hot spots. Furthermore, a thorough examination of neutron spectral moments reveals a close alignment with experimental data, quantitatively elucidating deviations from Maxwellian plasma behaviour observed in the experiments.

\textit{Model of Large-angle Collision} \label{sec:second}
To take large-angle collisions into account, the cut-off impact parameter \(b_c\) should be determined at first. Let \(\lambda_{\text{D}}\), \(n\), and \(q\) represent the Debye length of electrons, ion number density, and ion charge number, respectively. For a given target ion \(i\), when the binary potential \(\phi(r)=qe^{-{r}/{\lambda_{\text{D}}}}/(4\pi\epsilon_0r)\) of the incident ion equals the superposition potential of the background particles, the cut-off impact parameter \(b_\phi\) could be determined \cite{turrell2014effects}, i.e.,
\begin{equation}\label{equation:1}
	\phi_{ji}(r_{ij}=b_\phi)
	={\frac{q_sI_s}{4\pi\epsilon_0}\left( \frac{4\pi n_s}{3}\right)^{1/3}},
\end{equation}
where
\[
	I_s=\frac{3}{x}\left( k^{1/3}+\frac{1}{x}\right) \exp(-k^{1/3}x)+\sum_{m=1}^{k}{\frac{\exp(-m^{1/3}x)}{m^{1/3}}},
\]
\(x=(4\pi n_s/3)^{-1/3}/\lambda_{\text{D}}\), \(r_{ij}\) is the distance between \(i\) and \(j\), \(s\) denotes the ion species of higher number density between \(i\) and \(j\), 
and \(k\) is the number of ions for which the potential is calculated in a discrete manner.

Notably, this comparison should be conducted independently for different types of collisions. When calculating \(b_\phi\), the ion species of the background particles should match the one with higher number density between the target and incident ions, eliminating the need to consider \(q\) in solving Eq.\ (\ref{equation:1}) This enables optimization of simulation efficiency by tabulating \(b_\phi\) depending solely on density and Debye length (detailed in SM.\ I). 

Introducing ion relative motion requires the closest distance between collision trajectories, given by
\(r_{\min}=b_\bot+\sqrt{b^2+b_\bot^2}\), where \(b_\bot={q_1q_2}/{(4\pi\epsilon_0\mu u^2)}\) represents the impact parameter for vertical scattering, with \(u\) as the relative velocity and \(\mu={m_im_j}/{(m_i+m_j)}\) as the reduced mass of ions. Substituting \(b_\phi=r_{\min}\) into this expression gives the equation satisfied by \(b_c=b_u\) as
\begin{equation}\label{equation:2}
	b_\phi=r_{\min}=b_\bot+\sqrt{b_u^2+b_\bot^2}.
\end{equation}
Here, \(b_\phi \) is obtained using the static Coulomb potential by solving Eq.\ (\ref{equation:1}). By solving Eq.\ (\ref{equation:2}) involving \(r_{\min}\) and \(b_\phi\), the newly obtained value \(b_c=b_u\) combines both dynamic and static screened potentials of neighbouring ions. 

Simulating particle collisions, especially ion interactions, requires precise consideration of scattering angles and collision type effects. Independent with large-angle collisions, the cumulative small-angle collisions facilitating multi-body interactions are consistently utilized to generate a scattering angle \(\theta\), where the Coulomb logarithm is modified by \(b_u\) (as shown in Eq.\ (\ref{equation:A2}) in SM. II). To align with Maxwellian simulations or equivalent solutions from the Vlasov-Fokker-Planck equation \cite{atzeni2004physics}, the variance of \(\tan^2(\theta/2)\) is constrained bellow 0.02 as in Eq.\ (11) of ref.\ \cite{sentoku2008numerical}, ensuring sufficiently small average \(\theta\) per time step. For large-angle collisions with \(b<b_c=b_u\), an additional scattering angle \(\theta_L\) is superimposed. 

Using the differential scattering cross section, the cumulative density function \(\mathcal{C}_L(\theta)\) for large-angle collisions is calculated as
\[
	\mathcal{C}_L(\theta)
	=\int_{\theta_c}^{\theta}{\frac{1}{\sigma_L}\frac{{\rm d}\sigma}{{\rm d}\Omega'}{\rm d}\theta'}
	=1-\frac{b_\bot^2}{b_c^2}\cot^2{\frac{\theta}{2}},
\]
where \(\sigma_L=\pi b_c^2\) is the total large-angle cross section and \(\theta_c\) corresponds to \(b_c=b_u\). Differential scattering cross sections with screened Coulomb potentials do not affect the integration results.

The large scattering angle \(\theta_L\) following \(\mathcal{C}_L(\theta)\) is \(\theta_L=2\arctan{(b_\bot/b_c\sqrt{1-U})}\), where \(U\) is a uniform random number between 0 and 1. After generating and superimposing \(\theta_L\) in the center-of-mass (CoM) system, post-collision ion velocities are calculated to conclude the main part of the collision simulations. The validation of our model and implementation are detailed in SM.\ II-V.

Quantum effects in ``close'' electron collisions necessitate separate study \cite{Hansen1983Thermal, Glosli2008Molecular, Benedict2012Molecular, Graziani2012Large, turrell2015self}, while our large-angle model focuses on ion interactions, specifically the collision mechanism between \(\alpha\)-particles and fuel DT ions, known as ``alpha-particle knock-on (AKN)''. Collisions of neutral neutrons, such as ``neutron knock-on (NKN)'', exhibit a diminished cross-section and are not within our current simulation scope.

\begin{figure*}[htbp]
	\includegraphics[scale=0.34]{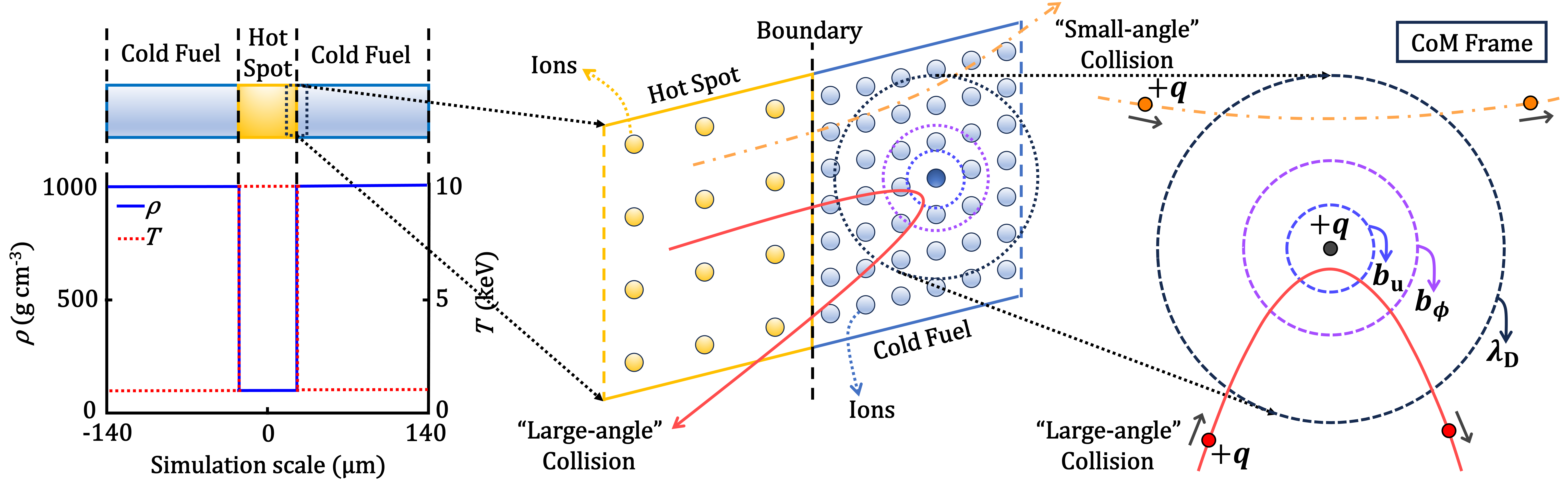}
	\caption{\label{fig:figure1} Initial condition schematic for simulations (left), with ``large-angle'' and ``small-angle'' ion collisions (middle) differing in impact parameters in CoM frame (right). Left: Hotspot initial temperatures and densities were ten times and one tenth of those of the initial cold fuels, respectively, matching the thermodynamically stable isobaric burning of the central ignition scheme. Middle and right: The center dark dot denotes the nucleus carrying charge \(+q\), the red solid or orange dotted line outlines the ion scattering trajectory for large or small angles, and dashed circles illustrate cross-section boundaries at impact parameter \(\lambda_{\text{D}}\), \(b_\phi\), \(b_u\) in colors from black, purple to blue. The area encompassed within the cutoff \(b_c=b_u\) determined by Eq. (\ref{equation:2}) serves as the cross section for large-angle collisions, whereas the opposite pertains to small-angle collisions.}
\end{figure*}

\textit{Simulation Results and Analysis}\label{sec:fourth} Employing the recently-developed hydrodynamic-kinetic hybrid code LAPINS \cite{wu2023headon}, which integrates ion kinetics and fusion reactions, we performed one-dimensional (1D) simulations of burning hot spots, with simulation details displayed in SM.\ VI. These simulations were initialized post-bang time (when the cold fuel shell is halted by the hot spot), incorporating a range of experimental initial hotspot conditions from NIF studies \cite{zylstra2019implosion, baker2020hotspot, hinkel2020optimization, thomas2020principal, doppner2020achieving, zylstra2021record, zylstra2022burning, zylstra2022experimental, kritcher2022design1, kritcher2022design2, baker2023reaching}. Despite the inherently three-dimensional (3D) nature of NIF implosions, recent experimental measurements and simulations indicate minimal asymmetric effects deviating from spherical symmetry, aligning the process with an ideal spherical implosion \cite{zylstra2019implosion, hinkel2020optimization, thomas2020principal, doppner2020achieving, zylstra2021record, kritcher2022design1, kritcher2022design2, baker2023reaching}. This validates the use of 1D simulations in Cartesian coordinates to capture the essential physics of high-energy ions and \(\alpha\)-particles in burning plasmas. 

\begin{figure}[htbp]
	\includegraphics[scale=0.5]{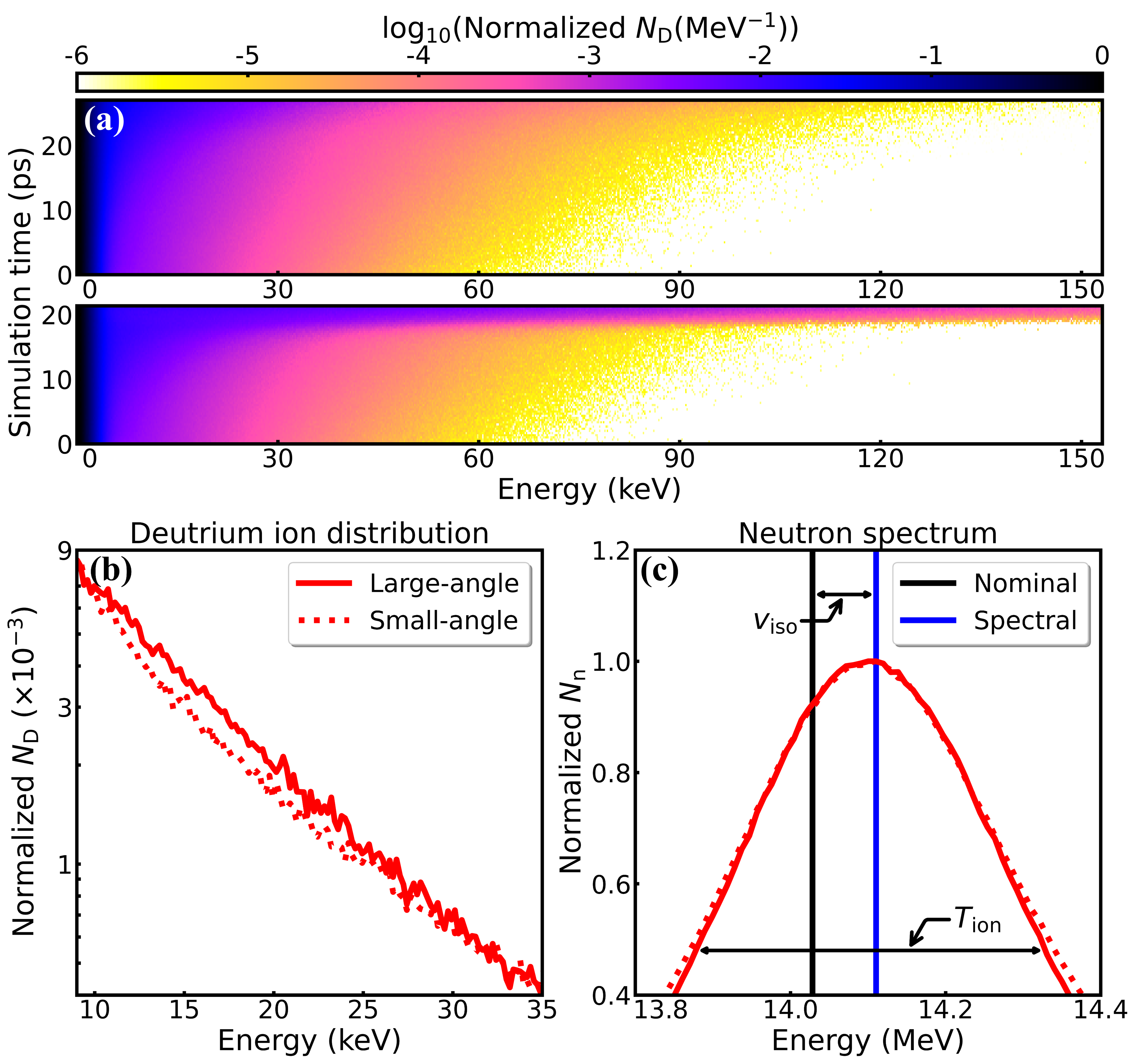}
	\caption{\label{fig:figure2} (a) Evolution of deuterium ion distribution \(N_\text{D}(t)\), where the color scale indicates the logarithm of \(N_\text{D}\); (b) Deuterium ion distribution \(N_\text{D}\) and (c) neutron spectra \(N_\text{n}\) from the moment before ignition (\(Y_\text{DT}\) close to \(3.09\times10^{16}\)). The image above and below in (a) or dashed and solid lines in (b, c) respectively represent the simulations before and after taking large-angle collisions into account. The ion isotropic velocity (\(v_\text{iso}\)) can be determined by the shift between the nominal neutron energy (the black vertical line in (c)) and the mean of \(N_\text{n}\) (blue), while the apparent ion temperature (\(T_\text{ion}\)) is calculated from the variance of \(N_\text{n}\), as shown in equation\ (\ref{equation:F2}) and (\ref{equation:F3}) of SM.\ VI.}
\end{figure}

\begin{figure}[htbp]
	\includegraphics[scale=0.5]{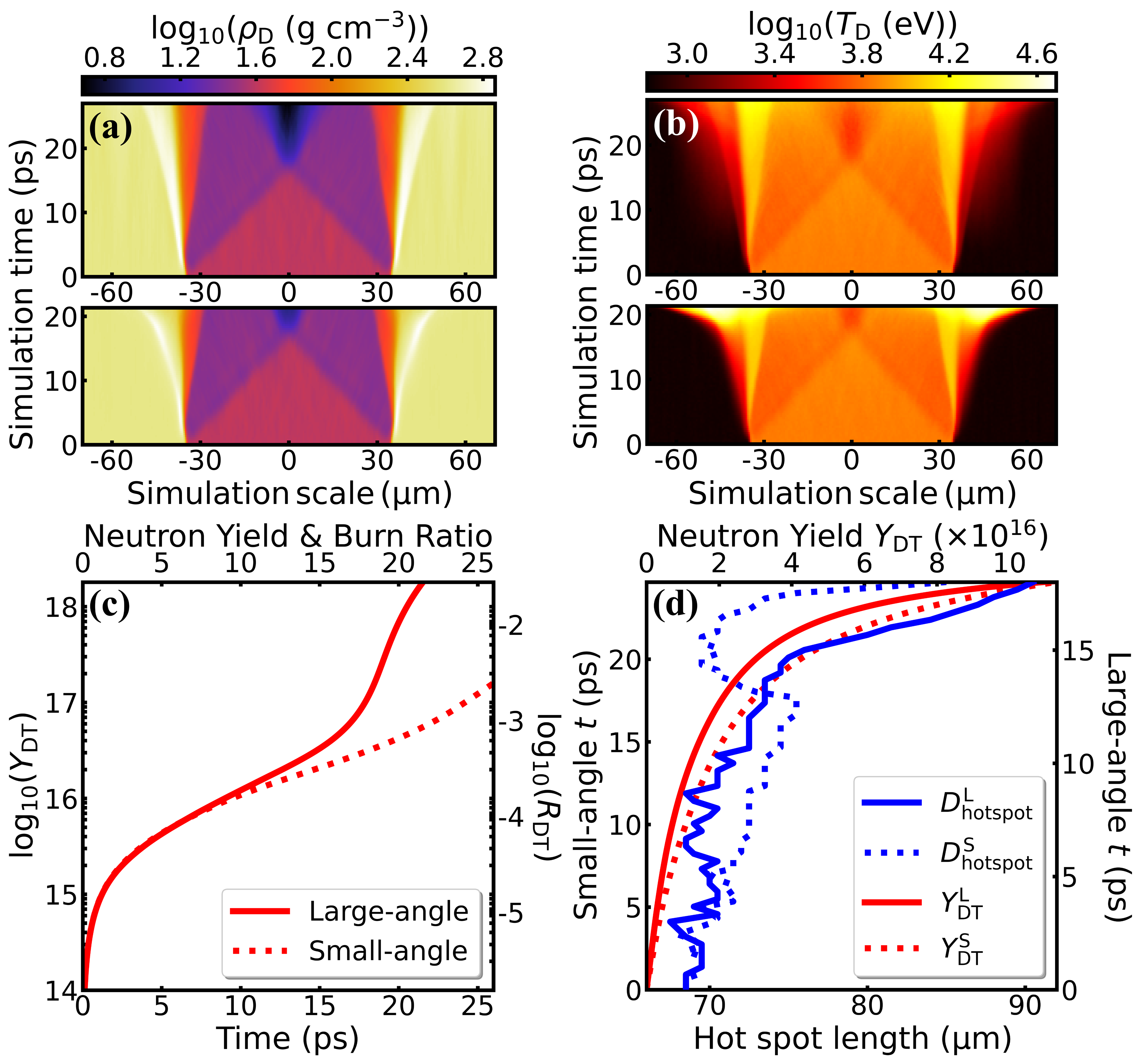}
	\caption{\label{fig:figure3} Evolution of (a) deuterium density \(\rho_\text{D}(t)\), (b) deuterium temperature \(T_\text{D}(t)\), (c) neutron yield \(Y_\text{DT}\) as well as burn ratio \(R_\text{DT}\) (logarithmic scale), and (d) hotspot length \(D_\text{hotspot}\) (at \(T_\text{D}>5.63\, {\rm keV}\)) with \(Y_\text{DT}\). The color scale in (a) and (b) indicates the density and temperature of deuterium, respectively. The image above and below in (a) and (b), dashed and solid lines in (c) and (d), or the superscripts ``S'' and ``L'' in (d) distinguish the simulations before and after considering large-angle collisions. In (d), the evolution of \(D_\text{hotspot}\) is indicated in blue, while \(Y_\text{DT}\) is indicated in red. The simulation time \(t\) in the longitudinal ordinates could serve as a hot spot burning time, scales of which are respectively adjust to compare the moment when the \(Y_\text{DT}\) approaches.
	}
\end{figure}

Figure \ref{fig:figure1} shows the schematic diagram of our simulation, the physical demonstration of ``large-angle” and ``small-angle” collisions with their impact parameters discussed in our model. Once the neutrons in the simulation were generated by DT fusions, they did not participate in scattering and were directly counted and accumulated in the simulation results, therefore it was also allowed to analyze the space and time integral neutron spectrum during hotspot burning. Figure\ \ref{fig:figure2}(a) includes a pair of simulations whose hot spots were initialized with \(7.0\, {\rm keV}\) in temperature, and 70 microns in length, which displays the evolution of deuterium ion distributions. In Fig.\(\ \)\ref{fig:figure2}(b) and (c), when the total \(Y_\text{DT}\) approaches \(3.09\times10^{16}\) (before ignition), the deuterium ion distributions and neutron spectra before and after considering the large-angle collisions are compared. Furthermore with the same simulation, figure \ref{fig:figure3} shows the density and temperature evolution of the hot spots in (a, b), as well as the \(Y_\text{DT}\) with corresponding DT fuels’ burn ratio \(R_\text{DT}\) in (c) that is increased by large-angle collisions. The length of the hot spot (with the deuterium ion temperature \(T_\text{D}>5.63\, {\rm keV}\)) over time is shown in Fig.\(\ \)\ref{fig:figure3}(d), where the \(Y_\text{DT}\) at the corresponding simulation time is also plotted. 


After considering large-angle collisions in simulations, notable disparities emerge in comparison to those simulations neglecting such collisions: Firstly, supra-thermal ions are observed below an energy threshold, revealing a substantial influence on the distribution profiles of low-energy ions. Secondly, alterations in neutron spectra are yielded, characterized by shifts in mean and variance and increases in \(Y_\text{DT}\). Thirdly, supra-thermal ion deposition are shaped and hotspot expansion were expedited, which may exhibit ``cooler'' burning congruent with NIF analyses \cite{hartouni2022evidence}. 

Regarding the first observation, in Fig.\(\ \)\ref{fig:figure2}(a), the logarithmic curve of the deuterium ion distributions in burning plasmas could be roughly segmented into two components: a low-temperature segment reflecting cold fuel and a high-temperature hotspot depicted in Fig.\(\ \)\ref{fig:figure2}(b), illustrating a supra-thermal ion tail with a \(\sim32\, {\rm keV}\) cut-off. Neutron analysis \cite{hartouni2022evidence} similarly hypothesizes that a Maxwellian distribution augmented by a supra-thermal tail (cut-off at \(35\, {\rm keV}\)) elucidates the interdependence between isotropic ion velocity (\(v_\text{iso}\)) and apparent ion spectral temperature (\(T_\text{ion}\)) beyond hydrodynamic boundaries. 
Accordingly, stemming from large-angle collisions, the heightened supra-thermal ion distribution reinforces experimental neutron analysis \cite{hartouni2022evidence, moore2023neutron}. In addition, supra-thermal ion distributions predicted by the large-angle collision model was verified by unprecedented MD simulations involving millions of particles and screened potentials for ion interactions (detailed in SM. V).

With respect to the second observation, large-angle collisions, similar to head-on encounters with ion relative velocities vastly exceeding the CoM velocity \cite{crilly2022constraints}, alter the neutron spectra profile in Fig.\(\ \)\ref{fig:figure2}(c), potentially elevating the relationship between \(v_\text{iso}\) and \(T_\text{ion}\) above Maxwellian predictions. Moreover, incorporating large-angle collisions, Fig.\(\ \)\ref{fig:figure3}(c) demonstrates either an increased \(Y_\text{DT}\) at a given time or an ignition advancement by \(\sim 10\, {\rm ps}\) for a certain \(Y_\text{DT}\). This is attributed to the amplification of fusion cross sections and output power resulting from supra-thermal ion depositions, particularly pronounced near hotspot boundaries. As evident in Fig.\(\ \)\ref{fig:figure3}(a) and \ref{fig:figure3}(b), simulations considering large-angle collisions exhibit a quicker burn wavefront traversing thicker fuel, attaining higher ion temperatures, densities and a more sufficient burn. 
Note that the Double-Cone Ignition (DCI) scheme \cite{zhang2020double} leveraging fast electrons to directly heat high-density fuel, could significantly capitalize on abundant supra-thermal ions from large-angle collision, thereby boosting neutron yields.

\begin{figure}[htbp]
	\includegraphics[scale=0.26]{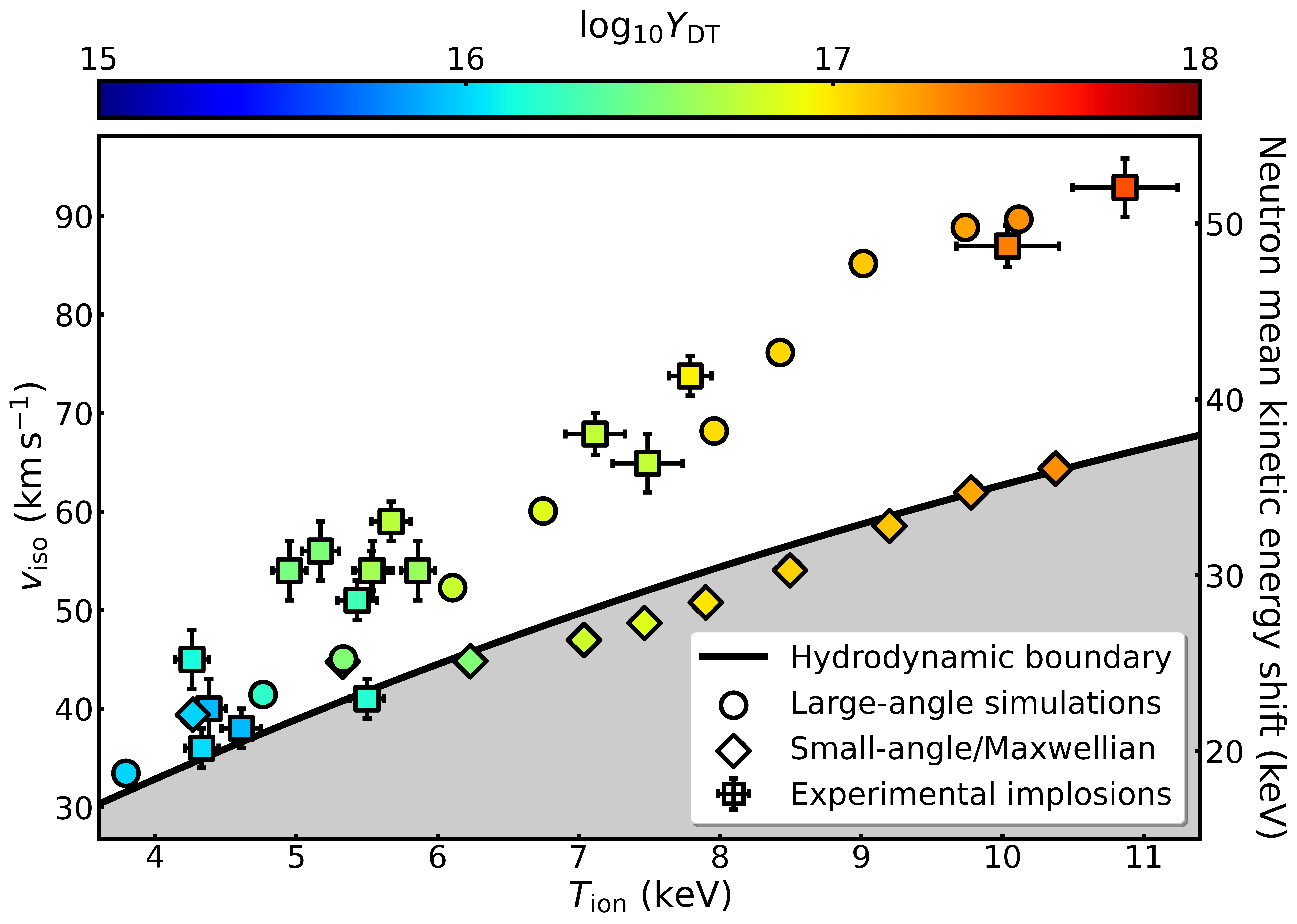}
	\caption{\label{fig:figure4}Interdependence of isotropic velocity (\(v_\text{iso}\)) and ion (spectral) temperature (\(T_\text{ion}\)). Diamonds and circles represent the simulation before and after considering large-angle collisions, respectively, while the experimental data represented in square symbols is from NIF (table 1 in ref. \cite{hartouni2022evidence} and figure 7 in ref. \cite{moore2023neutron}). It is “hydrodynamically accessible” \cite{hartouni2022evidence, moore2023neutron} for the gray shaded region, with an upper boundary represented by the thick black line. The color scale indicates the total neutron yield (\(Y_\text{DT}\)). Effects of bulk plasma velocity was removed for determining neutron mean kinetic energy shift (\(\Delta E=0.56037v_\text{iso}\), displayed on the right axis) and spectral temperature (Var(\(E_\text{n}\))), as shown in SM. VI}
\end{figure}

Pertaining to the third observation, large-angle collisions diminish the mean free path of thermal ions, mitigating the Knudsen layer effect that reduce the fusion reactivity due to tail fuel ions losses \cite{Molvig2012Knudsen, Albright2013Revised}, which facilitate the rapid and localized energy deposition of \(\alpha\)-particles, occasionally leading to “cooler” implosions. In Fig.\(\ \)\ref{fig:figure3}(b), within  \(21\, {\rm ps}\) of ignition, the expansion rate of hotspot boundaries accelerates from \(190\, {\rm km/s}\) to \(1190\, {\rm km/s}\) upon considering large-angle collisions, suggesting a nearly sixfold faster expansion. Furthermore, Fig.\(\ \)\ref{fig:figure3}(d) intuitively demonstrates that at the same \(Y_\text{DT}\), incorporating large-angle collisions increase the hotspot length \(D_\text{hotspot}\) (defined by \(T_\text{D}>5.63\, {\rm keV}\)) higher and earlier. Consequently, average ion temperatures decreases due to energy conservation, contributing to departures from Maxwellian behavior in spectral ion temperature below \(\sim7.5\, {\rm keV}\) in Fig.\(\ \)\ref{fig:figure4}. Additionally, in Maxwellian simulations (Fig.\(\ \)\ref{fig:figure3}(b)), the heated cold fuel surrounding the hot spot, with temperatures between \(3.1\, {\rm keV}\) and \(4.6\, {\rm keV}\), contributes negligibly to the burning process. However, in Maxwellian simulations above \(\sim7.5\, {\rm keV}\), the heating of cold fuels could potentially expand the hot spot, and elevate its temperature to approach simulations considering large-angle collisions at the same \(Y_\text{DT}\), aligning with Fig.\(\ \)\ref{fig:figure4}. Parenthetically, fuels with \(T_\text{D}<5.63\, {\rm keV}\) are excluded from \(D_\text{hotspot}\), accounting for its non-monotonic behaviour.

To validate our theory and simulations by experiments, we also leverage neutron analysis as a pivotal diagnostic tool. By incorporating both large-angle and small-angle collisions in LAPINS, we finally conduct 1D simulations to delved into the implications of ion kinetics on the \(v_\text{iso}\)-\(T_\text{ion}\) correlation. During neutron energy (\(E_\text{n}\)) analysis, \(v_\text{iso}\) and \(T_\text{ion}\) were determined from the spectral moments, reflecting shifts in mean or variance from nominal monoenergetic neutron energies (detailed in SM. VI). Fig.\(\ \)\ref{fig:figure4} compares the \(v_\text{iso}\)-\(T_\text{ion}\) relationship before and after considering large-angle collisions, focusing on a series of data pairs where \(Y_\text{DT}\) values approximate experimental measurements under identical initial conditions (hotspot temperature ranging from \(3.5\, {\rm keV}\) to \(10.0\, {\rm keV}\) with 70 microns in length). Experimental NIF data (table 1 in reference \cite{hartouni2022evidence} and figure 7 in reference \cite{moore2023neutron}) are also plotted for comparison.

Compared with Maxwellian simulations, our simulations incorporating large-angle collisions reveal a \(v_\text{iso}\)-\(T_\text{ion}\) relationship that aligns closely with experimental findings \cite{hartouni2022evidence, moore2023neutron}. As the plasma approaches sustained thermonuclear burn, mean energy upshifts (\(\Delta E\)) heighten, underscoring the growing importance of large-angle collisions in high temperatures. This enhancement arises from increased ion energy and \(\alpha\)-particle population, enabling ions at small spacing (or collision parameter) to penetrate Coulomb barriers during large-angle collisions. These collisions resemble head-on collisions which are characterized by high ion relative velocities and low CoM velocities. Correspondingly, via energy conservation, high ion relative velocities isotopically amplify neutron velocities during fusion reactions, while low CoM velocities mitigate Doppler broadening of spectral temperatures, resulting in boosted mean neutron kinetic energies and narrowed spectral temperatures \cite{crilly2022constraints}. Therefore, our simulations demonstrate that large-angle collisions could deviate \(v_\text{iso}\)-\(T_\text{ion}\) relationship from Maxwellian or small-angle predictions, accurately capturing experimental observations in burning plasmas.

Worth mentioning, the limit range of the \(v_\text{iso}\)-\(T_\text{ion}\) relationship derived from fully symmetric isotropic ion distributions has been established \cite{crilly2022constraints}. Current spectral moment analysis for spectral temperatures above \(10\, {\rm keV}\) still aligns with this theory \cite{crilly2022constraints}. Large-angle collisions, considered from the perspective of CoM motion, embody an isotropic mechanism and should also leading to isotropic ion distributions. Consequently, the congruency between experimental analysis and simulation results embedding large-angle collisions in the \(v_\text{iso}\)-\(T_\text{ion}\) relationship confirms this theory through kinetic simulations.
\[\]

\textit{Conclusions}\label{sec:fifth}
This Letter has elaborated on the effects of large-angle collisions in ICF burning plasmas. A model capturing ion kinetics in binary Coulomb collisions is shown to be pivotal for accurately modelling both the large-angle collisions and small-angle collisions. Simulation results indicate that the consideration of large-angle collisions could confirm the supra-thermal ion distribution below an energy cutoff in experimental neutron analysis, offering insights into their generation. Furthermore, it has significant impacts on promoting the ignition moment by \(\sim 10\, {\rm ps}\) and enhancing \(\alpha\)-particle deposition in and around the hot spot, resulting in a sixfold faster expansion. Finally, as experiments approach burning plasma states, the observed increasing departure from Maxwellian or hydrodynamic behaviour could be reliably explained and predicted by accounting for large-angle collisions. This work advances our understanding of burn propagation and ignition requirements in ICF plasmas, providing a valuable tool for future research on kinetic effects in burning and ignited plasmas, with potential to guide the design and improvement of ignition schemes.

This work was supported by the Strategic Priority Research Program of Chinese Academy of Sciences (Grant No. XDA250010100 and XDA250050500), National Natural Science Foundation of China (Grant No. 12075204) and Shanghai Municipal Science and Technology Key Project (Grant No. 22JC1401500). D. Wu thanks the sponsorship from Yangyang Development Fund.

\bibliography{Reference}

\clearpage
\setcounter{figure}{0}
\renewcommand{\thefigure}{A\arabic{figure}}
\renewcommand{\thetable}{A\arabic{table}}
\begin{center}
	\textbf{SUPPLEMENT MATERIALS}
\end{center}

\section{I. Optimization of Simulation Efficiency}
\begin{figure}[htbp]
	\includegraphics[scale=0.35]{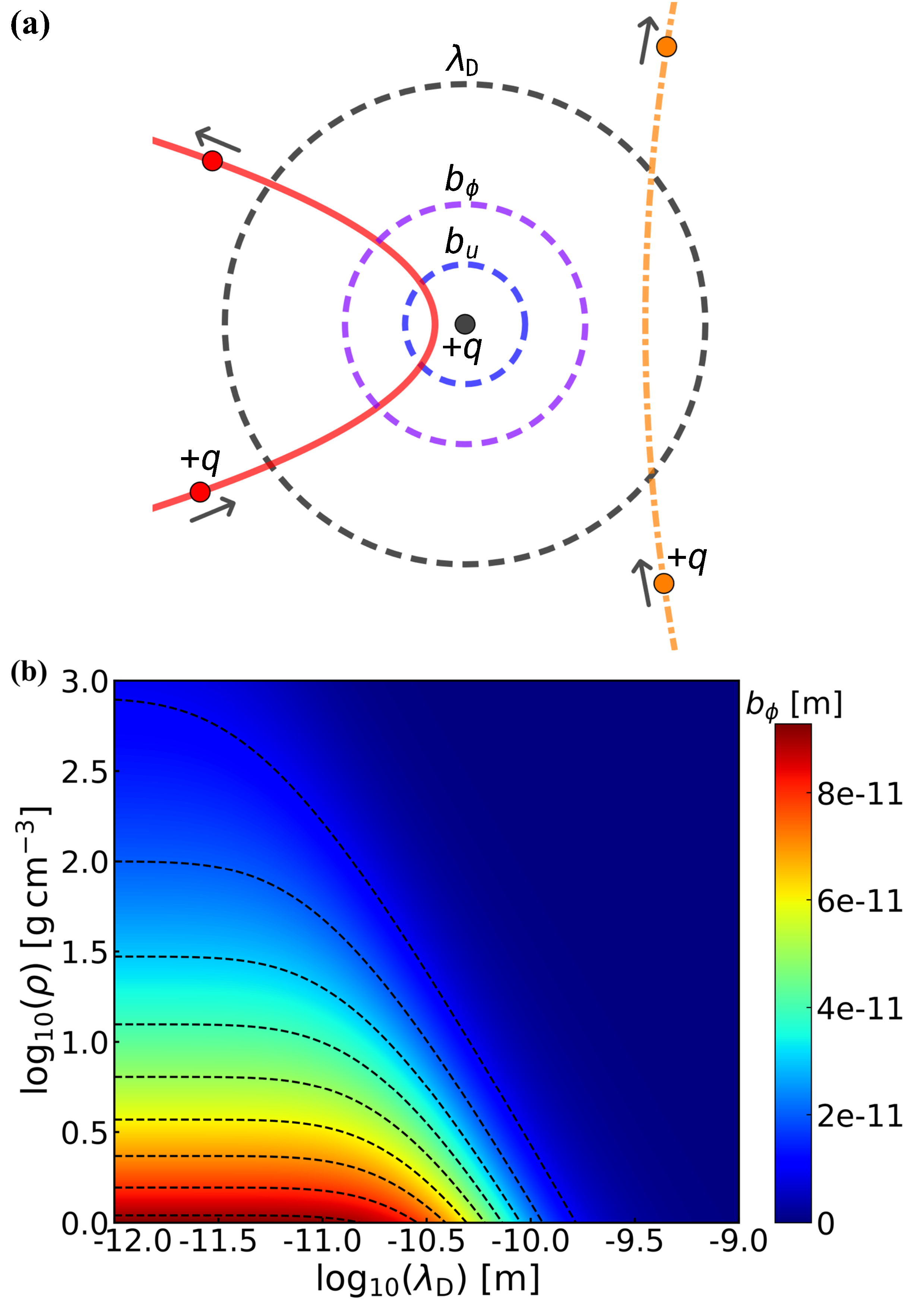}
	\caption{\label{fig:figureS1} (a) Schematic diagram of impact parameters and collision cross sections in the CoM frame. Dashed circles in black, purple, and blue represent boundaries of cross sections with radius at impact parameter \(\lambda_D\), \(b_\phi\), and \(b_u\). The center dot denotes the \(+q\) charged nucleus, red solid and orange dotted lines illustrate scattering trajectories for large and small angle collisions, respectively. The area within \(b_c=b_u\) (determined by Eq. (\ref{equation:2})) in LAPINS calling \(b_\phi\) from a table based on the static screened potential \cite{turrell2014effects} is the cross section of large-angle collisions, while the exterior corresponds to small-angle collisions; (b) Values of \(b_\phi\) in the loaded table for varying electron Debye lengths \(\lambda_{\text{D}}\) and DT ion density \(\rho\). The color scale indicates magnitudes of \(b_\phi\) and the black dashed line outlines the contour of \(b_\phi\).}
\end{figure}

It was time-consuming to solve Eq.\ (\ref{equation:1}) for cut-off \(b_c\) if the module of large-angle collisions was directly embedded into the code LAPINS. To optimize the simulation efficiency, the values of \(b_\phi\) in the range of burning plasma conditions of interest were pre-calculated in the tabular form and loaded into the code. During the simulation, the \(b_\phi\) table would be interpolated using the reciprocal of the distances between \(b_\phi\) points as weights when \(b_\phi\) is called to calculate the new cut-off using Eq.\ (\ref{equation:2}). The values of \(b_u\) are not pre-input because \(b_\phi\) only depends on the DT ion density \(\rho\) and the electron Debye length \(\lambda_{\text{D}}\), while \(b_u\) also depends on the relative velocity of binary ions in a single collision. Usually, \(\lambda_D>b_\phi>b_u\). The relative relationship between each impact parameter is schematically shown in Fig.\(\ \)\ref{fig:figureS1}(a), and Fig.\(\ \)\ref{fig:figureS1}(b) shows the table of \(b_\phi\) loaded into the code for varying \(\rho\) and \(\lambda_{\text{D}}\).

The code is capable of calling \(b_\phi\) from the table without knowing the ion species, which comes from the assumption that the ions are uniformly distributed in local \cite{turrell2014effects}. For this reason, the only variable required for calculating the potential is the ion number density, which can be obtained from the \(\rho\) axis of the table. Moreover, the ion charge \(q\) could be eliminated during the solution of Eq. (\ref{equation:1}) in the main text, because the comparison of the screened potential in a single collision should only occur between ions of the same species \(s\). Furthermore, we choose \(s\) as the ion species with the higher density in the binary collision to improve the stability of rare ions participating in large angle collisions. As a consequence, calling \(b_\phi\) does not involve variables related to the ion species.

\section{II. Modified Coulomb logarithm}
The variance of scattering angles in small-angle collision algorithm is related to the Coulomb logarithm, which could be modified \cite{wu2020particle} by the screened Coulomb potentialand the previously \(b_c\) calculated in the main text. Under the Born approximation and with a screened potential \(\exp(-r/\lambda_{\text{D}})/r\), the differential scattering cross section is \cite{zheng2009plasma}
\begin{equation}\label{equation:A1}
	\sigma(\theta)\sim\left[ \sin^2\left( \frac{\theta}{2}\right)+\epsilon^2 \right]^{-2},
\end{equation}
where
\[
\epsilon=
\begin{cases}
	{h}/{(2\mu u\lambda_{\text{D}})},	& {u}/{c}>{q_1q_2}/{(2\pi\epsilon_0hc)} \\ 
	{b_\bot}/{\lambda_{\text{D}}}, 	& {u}/{c}<{q_1q_2}/{(2\pi\epsilon_0hc)} \\ 
\end{cases}
\]
and \(c\) is the speed of light. Substituting Eq.\ (\ref{equation:A1}) into the Coulomb logarithm \(\ln{\Lambda}\sim\int_{0}^{\theta_c}{\sin{\theta}\sin^2(\theta/2)\sigma(\theta){\rm d}\theta}\) gives
\begin{equation}\label{equation:A2}
	\ln{\Lambda}
	\sim\ln{\left( \frac{1-\cos{\theta_c+2\epsilon^2}}{2\epsilon^2}\right)} -\frac{1-\cos{\theta_c}}{1-\cos{\theta_c}+2\epsilon^2},
\end{equation}
where \(\tan{(\theta_c/2)}=b_\bot/b_c\), and the \(\ln{\Lambda}\) is used to generate the scattering angles in the small-angle collision method \cite{takizuka1977binary, nanbu1997theory}.
Besides, the electron Debye length \(\lambda_{\text{D}}\) in the simulation is modified to the larger of \(\lambda_{\text{D}}\) and the spacing between ions, while the impact parameter \(b_\bot\) for the vertical scattering is modified to the larger of \(b_\bot\) and the de Broglie wavelength \cite{wu2020particle}.

\section{III. Verification of knock-on ions}
Validity of large-angle collision simulation for a specific \(b_c\) could be verified by a benchmark \cite{turrell2015self} based on the theoretical generation rate of knock-on ions \cite{helander1993formation}. In the simulation, \(\alpha\)-particles satisfy a mono-energetic distribution with a kinetic energy of \(E_\alpha=3.54\, {\rm MeV}\) and deuterium ions are barely cold. The number densities of \(\alpha\)-particles and deuterium ions are equal as \(n_\alpha=n_{\text{D}}=5\times10^{31}\, {\rm m^{-3}}\), and the cut-off of impact parameter is intentionally chosen as \(b_c=b_\bot\).

The equation for yield rate \(Q_{\text{D}}\) of the knock-on deuterium ions per unit volume per unit time per unit energy \cite{turrell2015self, helander1993formation} is
\begin{equation}\label{equation:C1}
	Q_{\text{D}}{\rm d}E_{\text{D}}
	=\left( \frac{q_{\text{D}}q_\alpha}{4\pi\epsilon_0}\right) ^2\frac{2\pi n_{\text{D}}n_\alpha}{m_{\text{D}}E_{\text{D}}^2\sqrt{2E_\alpha/m_\alpha}}{\rm d}E_{\text{D}},
\end{equation}
where \(q_{\text{D}}\) and \(q_\alpha\) denote the charges of deuterium ions and \(\alpha\)-particles. 
Since the impact parameter \(b\) takes values from \(0\) to \(b_\bot\), the kinetic energy of knock-on ions is bounded with \(E_{\max}=4E_\alpha m_{\text{D}} m_\alpha/(m_{\text{D}}+m_\alpha )^2\) and \(E_{\min}=E_{\max}/2\) \cite{turrell2015self}.

\begin{figure}[htbp]
	\includegraphics[scale=0.35]{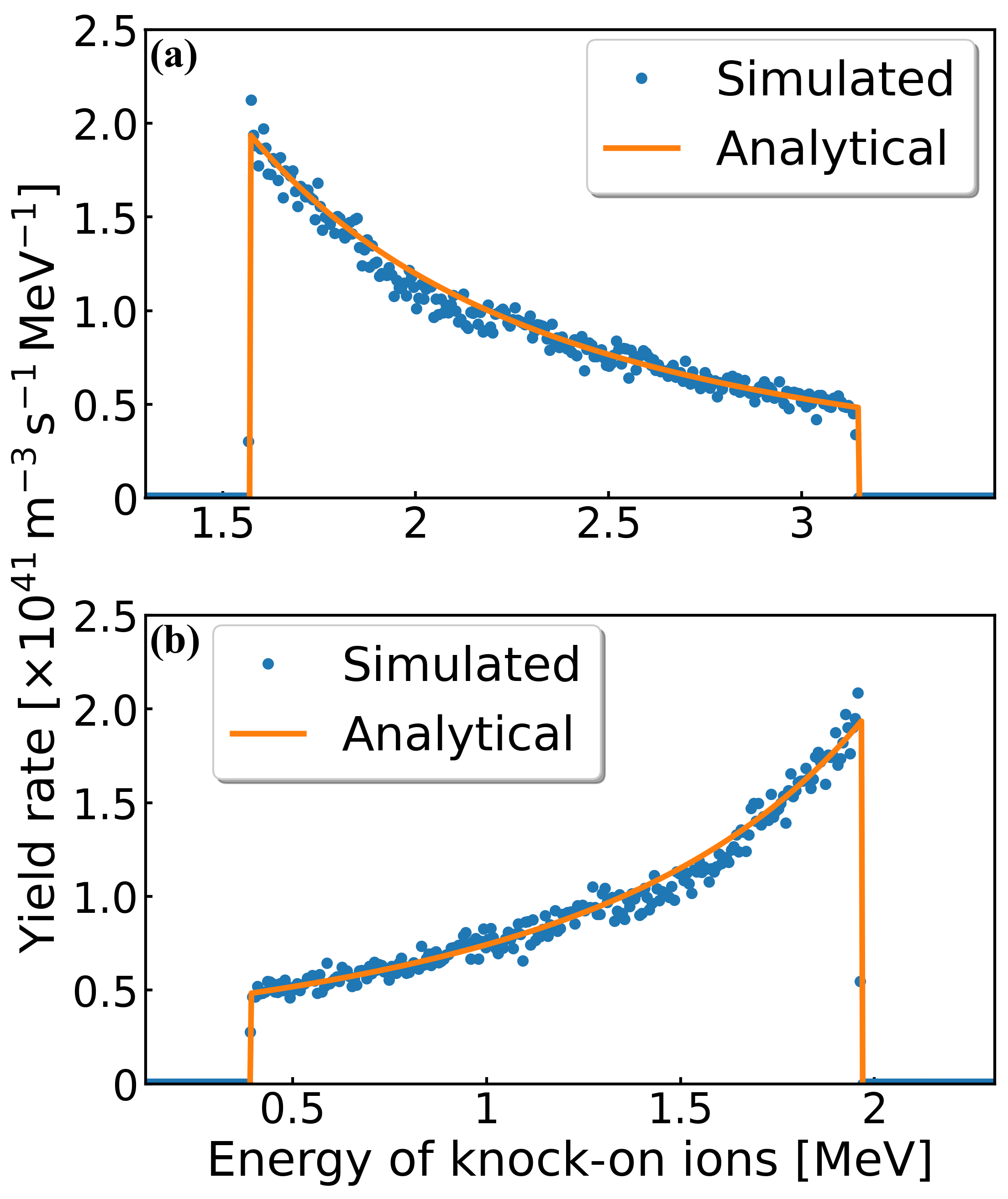}
	\caption{\label{fig:figureS2} Plot of yield rate when \(\alpha\)-particles collide with cold deuterium ions. The blue dots are simulation data and the red curves are the theoretical yield rates calculated by Eq. (\ref{equation:C1}). (a) Yield rate of knock-on deuterium ions; (b) Yield rate of \(\alpha\)-particles after large-angle collisions with the deuterium ions.}
\end{figure}

To verify our simulations against the theoretical rate, only large-angle collisions between \(\alpha\)-particles and deuterium ions are simulated. Moreover, only the first generation of knock-on ions are simulated, which means that exactly one collision step is simulated on the LAPINS code so that a single large-angle collision occurs between each pair of deuterium ions and \(\alpha\)-particles. Additionally, the time step is restricted small enough for \(\alpha\)-particles to change small energies in collisions, thus allowing \(\alpha\)-particles to fit the mono-energetic distribution in a close manner.

The yield rate of knock-on deuterium ions obtained by analyzing the energy distribution is shown in Fig.\(\ \)\ref{fig:figureS2}(a), while that of \(\alpha\)-particles that lose energies in collisions with cold deuterium ions could be calculated similarly, noting that the total energy is equal to the initial kinetic energy \(E_\alpha=3.54\, {\rm MeV}\). The curve of the \(\alpha\)-particles in Fig.\(\ \)\ref{fig:figureS2}(b) naturally follows a symmetrical shape with respect to the curve of the deuterium ions in Fig.\(\ \)\ref{fig:figureS2}(a). 

The simulation data points essentially conform to the shape of the theoretical curve, indicating that the modelling of large-angle collision is as expected. As a result, the principles and mechanisms of the original large-angle collision simulation method \cite{turrell2015self} could be followed in LAPINS, while the value of \(b_c\) could be altered.

\section{IV. Verification of temperature equilibrium}
\begin{figure}[htbp]
	\includegraphics[scale=0.35]{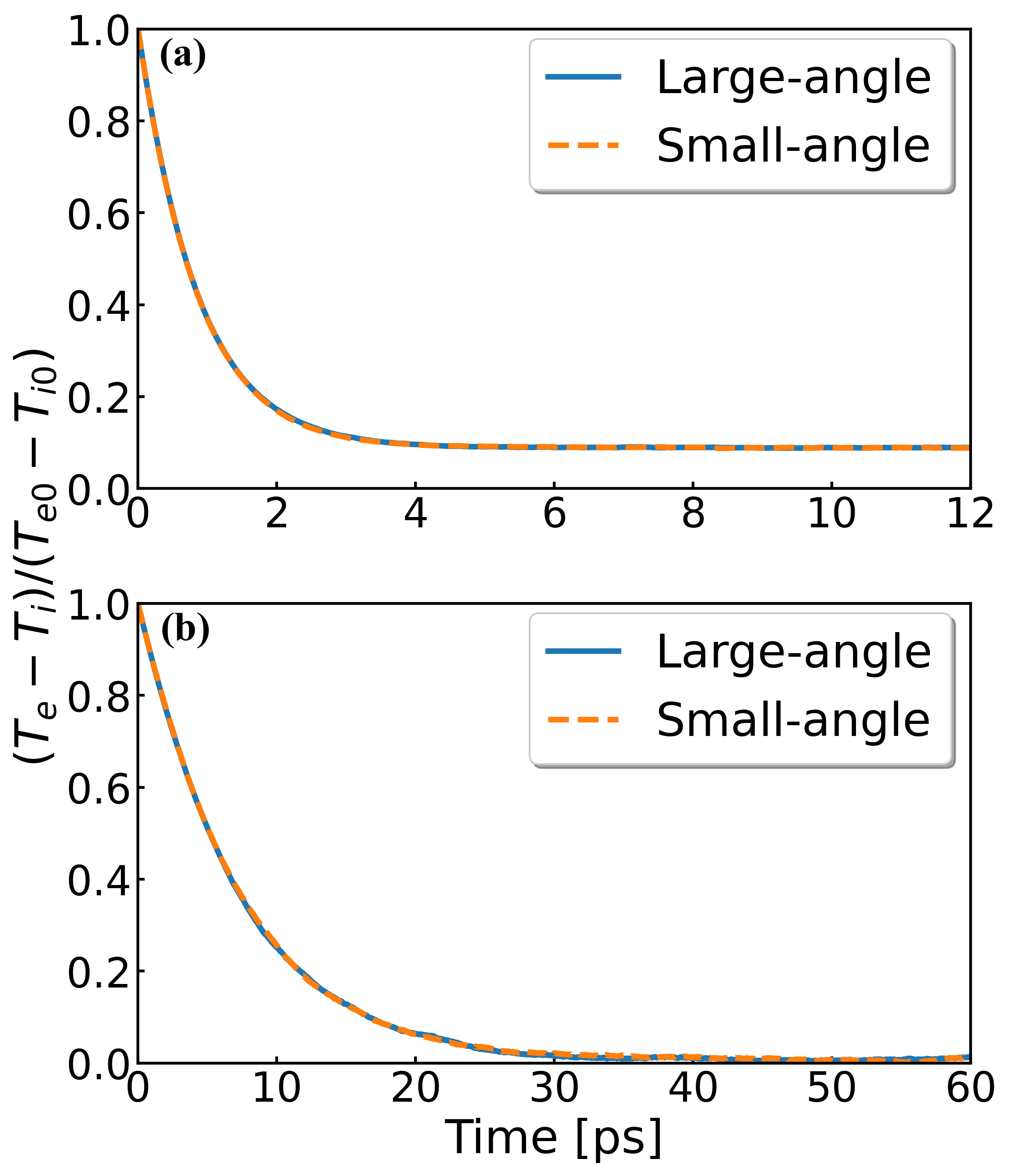}
	\caption{\label{fig:figureS3} The rate of temperature equilibrium as a function of time. The red line is when only the cumulative small-angle collisions are included, and the blue line is when taking large-angle collisions into account. (a) The initial temperature of deuterium ion is \(T_{i0}=66.8\, {\rm eV}\) and that of electron is \(T_{e0}=533\, {\rm eV}\); (b) The initial temperature of deuterium ion is \(T_{i0}=6.66\, {\rm keV}\) and that of electron is \(T_{e0}=10.0\, {\rm keV}\).}
\end{figure}

The temperature equilibrium of deuterium-electron was simulated at initial temperatures of \(T_{\text{D}}=66.8\, {\rm eV}\) and \(T_{e}=533\, {\rm eV}\) or \(T_{\text{D}}=6.66\, {\rm keV}\) and \(T_e=10.0\, {\rm keV}\) with a fixed ion density of \(\rho_{\text{D}}=83.6\, {\rm g/cm^3}\). For neutralization, the number density of electrons was set to be the same as deuterium ions. The rate of temperature equilibrium is defined as the ratio of the ion-electron temperature difference to the initial temperature difference, \((T_e-T_i)/(T_{e0}-T_{i0})\), whose variation with time is shown in Fig.\(\ \)\ref{fig:figureS3}.

In the benchmark of temperature equilibrium, the cut-off is \(b_c=b_u\). We also neglect energy losses and heating processes such as bremsstrahlung losses and fusion reactions. 
Under the condition in Fig.\(\ \)\ref{fig:figureS3}(a), the temperature equilibrium rate is not equal to zero even in the end, which may result from the temperature being so low that the quantum degeneracy leads to the final temperature difference. Under quantum degenerate state, the electron temperature is no longer a pure function of averaged kinetic energy, however it also depends on density. The quantity $T_e$ in Fig.\(\ \)\ref{fig:figureS3}(a) is actually the averaged electron kinetic energy, containing the contribution from the Fermi energy.
For the simulations in Fig.\(\ \)\ref{fig:figureS3}(b), the quantum degeneracy could be ignored, therefore the temperature equilibrium rate is almost zero when the curve is stable.

From the curves in Fig.\(\ \)\ref{fig:figureS3}, we see the effect of large-angle collision on the electron-ion relation rate is negligible regardless of whether the quantum degeneracy can be neglected or not. 
Simulations indirectly indicate that small-angle collisions still dominate in the process of thermal relaxation even though the large-angle collision can exchange more energy in a single collision. 
Small-angle collisions contribute the energy exchange for the majority of particles, thereby dominating the temperature equilibrium.

\section{V. Verification by the MD simulation}
Molecular dynamics (MD) could include a comprehensive range of Coulomb collisions across various angles naturally \cite{Graziani2012Large} to delve into the ion kinetics. To verify the accuracy of the supra-thermal ion distribution generated by our large-angle collisions model, a simplified inertial confinement fusion hotspot was performed by both a MD simulation and our LAPINS code, with reference to Figure 5 of ref. \cite{turrell2015self}. 

\begin{figure}[htbp]
	\includegraphics[scale=0.27]{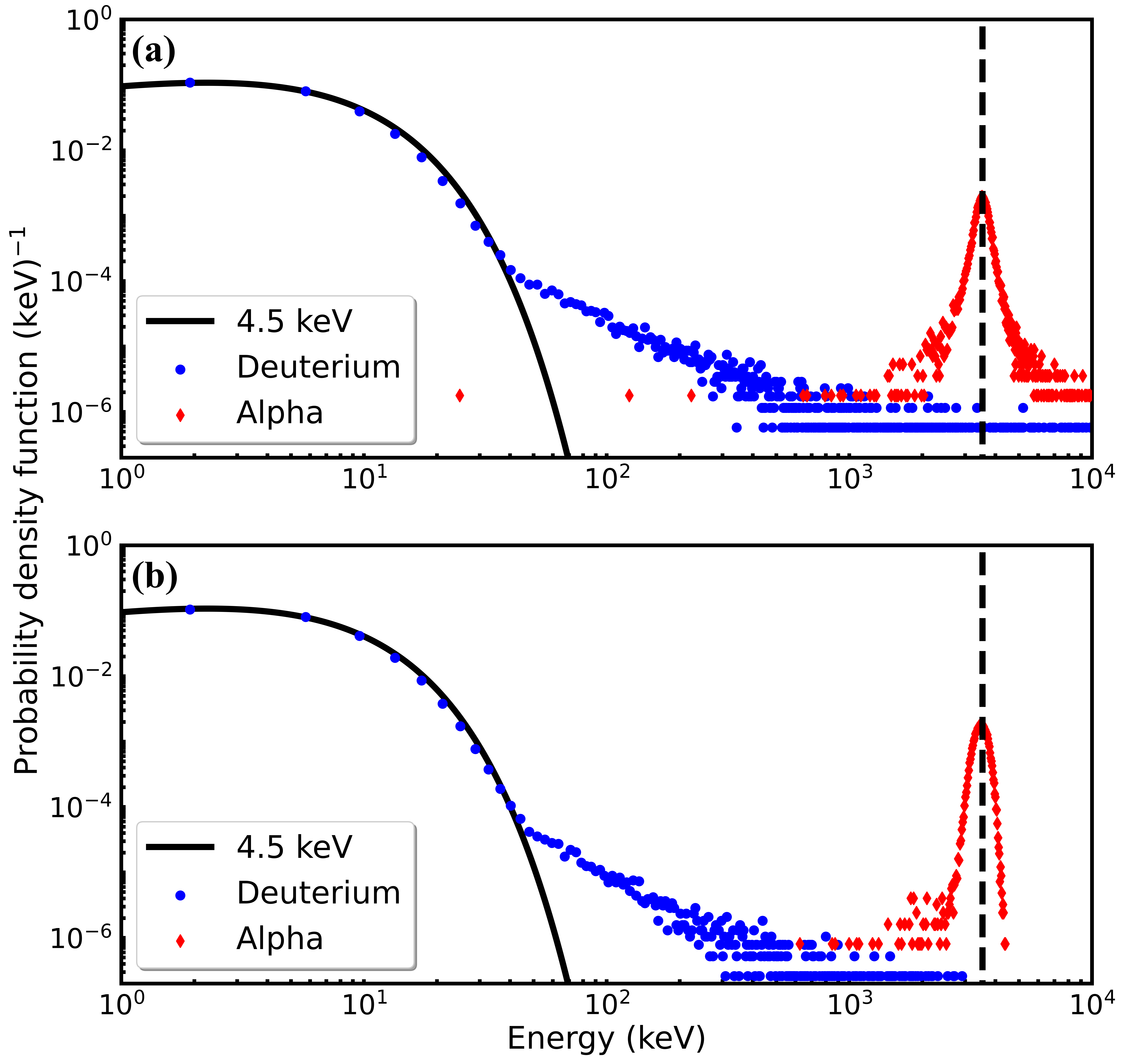}
	\caption{\label{fig:figureS4} Deuterium and \(\alpha\) ion distribution in simulations from (a) MD at \(220\, {\rm fs}\); (b) LAPINS at \(400\, {\rm fs}\). The black solid line represents the fitted Maxwell distribution at \(4.5\, {\rm keV}\), the blue dots and the red diamonds represent the deuterium and \(\alpha\)  ion distribution, respectively. The initial energy \(E=3.54\, {\rm MeV}\) of \(\alpha\) is displayed as a dashed line.}
\end{figure}

The MD simulation was performed with a fully parallel code using the self-coded particle-particle particle-mesh (PPPM) method \cite{dharuman2017generalized} with the screened Coulomb interaction in the canonical ensemble, which was applied to deuterium-alpha mixtures with a total of 474600 ions in a cubic cell with periodic boundary conditions. The system contained deuterium initiated in thermal equilibrium at \(T=3\, {\rm keV}\) and \(n_\text{D}=4.52\times10^{31}\, {\rm m}^{-3}\) and 5\% \(\alpha\)-particles were added of \(E=3.54\, {\rm MeV}\). To ensure charge neutrality the particles are immersed in a uniform background of electrons. A same initialization was applied in the simulation performed by our LAPINS code, except that a total of one million deuterium, 50 thousand \(\alpha\)-particles and correspondingly 1.1 million electrons were uniformly initialized in 50 cells.

The time step in the MD simulation was chosen to be \(\Delta t=5×10^{-7}\, {\rm fs}\) to ensure good energy conservation (\(\Delta E/E<10^{-5}\)), while in our LAPINS code, \(\Delta t=1/3\, {\rm fs}\). The comparison of deuterium ion distribution from two method, as well as their fitted Maxwell distribution are shown in Fig.\(\ \)\ref{fig:figureS4}. It could be observed that in probability density function, the departure of Maxwellian behavior begins at \(10^{-4}\sim10^{-5}\) and continues at lower values. Moreover, although the simulation moment selected for comparison differs, the shape of deuterium ion distribution in each method resembles, which suggests that our large-angle model could generated a similar group of supra-thermal ion distribution to the method of MD.

\section{VI. Details of simulations and analysis}
For the moment of hot spot burning after the bang time, the laser has already been turned off and the laser energy has been stored in the hot spot in the form of thermal energy of ion population. Therefore, these simulations started from the formation of a burning plasma hot spot where the deposition energy of \(\alpha\)-particles exceeded the work done by the compression of internal explosion. Recent experimental measurements and simulations have shown that the symmetry control of hot spots in NIF has been continuously enhanced in recent years \cite{zylstra2019implosion, hinkel2020optimization, thomas2020principal, doppner2020achieving, zylstra2021record, kritcher2022design1, kritcher2022design2, clery2022explosion, baker2023reaching}, approaching the states of highly spherical symmetric hot spots suitable for one-dimensional (1D) simulations. Therefore, we perform 1D simulations in 560 uniformly sized cells with a total length of 280 microns in all the initial conditions, which includes an initial central hot spot with DT mass density \(\rho_\text{DT}=100\, {\rm g/cm}^{3}\) and cold DT fuels in the remaining areas with \(\rho_\text{DT}=1000\, {\rm g/cm}^{3}\), similar to experimental conditions at NIF \cite{Frenje2013Diagnosing, zylstra2019implosion, baker2020hotspot, hinkel2020optimization, doppner2020achieving, zylstra2021record, zylstra2022burning, zylstra2022experimental, kritcher2022design1, kritcher2022design2, baker2023reaching}. Each cell had 2000 quasi-particles for electrons, deuterium and tritium respectively, while the number of quasi-\(\alpha\)-particles produced by fusion is limited to 4000 per cell using agnostic conservative down-sampling for optimization \cite{Frenje2013Diagnosing}. All simulations kept DT number density the same, \(n_\text{D}=n_\text{T}\), and only DT and DD fusion reactions were included, excluding other minor fusion reactions such as TT.

\begin{figure}[htbp]
	\includegraphics[scale=0.34]{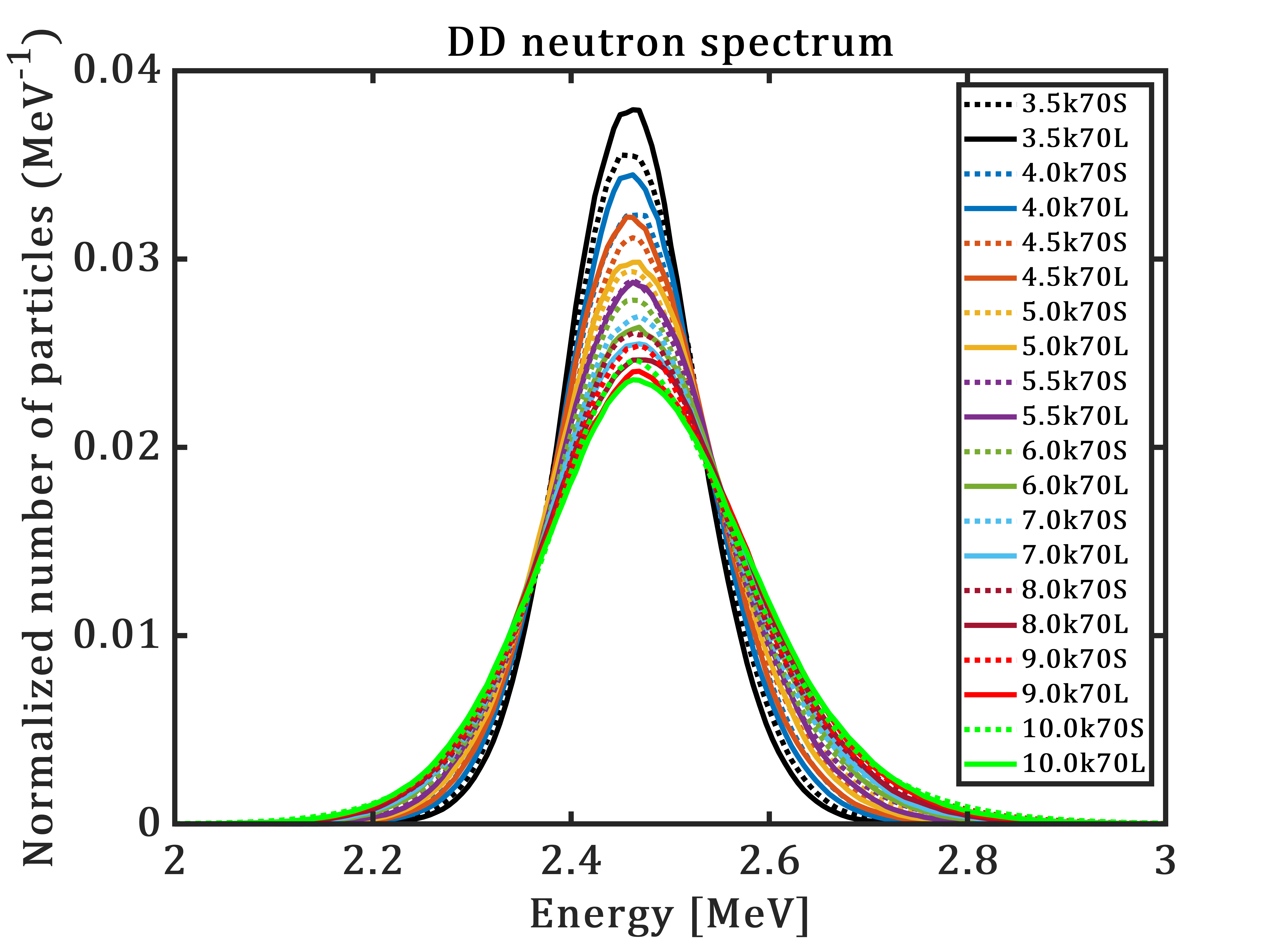}
	\caption{\label{fig:figureS6} The neutron spectrum from DD reactions produced at the same moment as in Table \ref{tab:tableS1}. Solid and dotted lines or ``L'' and ``S'' in the legend indicate simulations before and after considering large-angle collisions, and different colors of lines represent different LAPINS simulations.}
\end{figure}

The massive number of quasi-particles included in the simulation enables us to obtain neutron spectrum from DD reactions together with DT neutron spectrum, as shown in Figure \ref{fig:figureS6}. The width of the DD neutron spectrum in Fig.\(\ \)\ref{fig:figureS6} also represents the ion temperature. It can also be observed that when \(T_\text{init}<5.5\, {\rm keV}\), the ion temperature is lower in the simulation after considering the large-angle collisions, and the opposite is true when \(T_\text{init}>5.5\, {\rm keV}\), which is close to the results of DT neutron analysis.

A series of 1D simulations similar to historical explosions at NIF were performed in the hybrid-particle-in-cell code LAPINS. The initial conditions of these simulations differed in the initial plasma temperature and initial hot spot length, as shown in the first two columns of Table \ref{tab:tableS1}. Based on the simulations and data measured in the experiment at NIF \cite{baker2020hotspot, zylstra2022burning, zylstra2022experimental, kritcher2022design1, kritcher2022design2}, the length (\(D_\text{init}\)) of the initial hot spot was selected as 50 microns (indicated in the main text), with a temperature (\(T_\text{init}\)) range from \(4.64\, {\rm keV}\) to \(10.57\, {\rm keV}\), while all the initial temperatures of the initial cold fuels were set at one tenth of the corresponding initial hot spots. To make the initial hot spots of simulations consistent with the thermodynamically stable isobaric burning process, of the hot spot after the bang time in experiments of the central ignition scheme, the initial hot spots’ temperatures and densities were ten times and one tenth of those of the initial cold fuels, respectively.

For other simulation settings, the computational time step was \(1/3\, {\rm fs}\), the electromagnetic field on particles was considered, and energy non-conservation rates of these simulations were all below 1\%. A major loss mechanism for burning plasmas, bremsstrahlung from electron-baring ion collisions, was also taken into account in the simulation, whose rate is calculated as
\begin{equation}\label{equation:F1}
	W=1.69\times10^{-32}n_e\sqrt{T_e}\sum_{a}{n_aZ_a^2}\, [{\rm Wcm^{-3}}],
\end{equation}
where \(T_e\) is in \({\rm eV}\), \(n\) is in \({\rm cm^{-3}}\) and \(a\) runs over all baring ion species with ion atomic number of \(Z_a\) \cite{atzeni2004physics, sadler2019kinetic}.

In the neutron analysis of the main text, the relationship between isotropic ion velocity (\(v_\text{iso}\)) and anisotropic apparent ion temperature (\(T_\text{ion}\)) were determined \cite{hartouni2022evidence, munro2016interpreting} from the shift between the nominal neutron energy and the neutron spectra (\(E_\text{n}\)), using equations (\ref{equation:F2}) and (\ref{equation:F3}) from reference \cite{hartouni2022evidence}, i.e,
\begin{align}
	&\text{Var}(E_\text{n})\approx0.31401\text{Var}(u_\parallel)+6024.6\left\langle T_\text{ion} \right\rangle,\label{equation:F2}\\
	&\left\langle E_\text{n} \right\rangle\approx14.02839\text{MeV}+0.00056\left[ \left\langle u_\parallel \right\rangle + v_\text{iso} \right],\label{equation:F3}
\end{align}
where ``\(\text{Var}()\)'' and symbol ``\(\left\langle \cdot \right\rangle \)'' means the variance and the mean value over the space and time of variables, respectively. \(u_\parallel\) is the projection of bulk or average plasma velocity along the measurement direction, which is approximately sampled from the spatial and temporal distribution of fluid flow velocity \(U(x,t)\) in a simulation cell with temperature related weight \(w\left[ t,T(x,t)\right]\), i.e.,
\begin{align}
	&u_\parallel(x,t)=\frac{w\left[ t,T(x,t)\right] U(x,t)}{\underset{x}{\max}\left\lbrace w\left[ t,T(x,t)\right] \right\rbrace },\label{equation:F4}\\
	&w\left[ t,T(x,t)\right]=\left[ T_\text{h}(x,t)T_\text{e}(x,t)\right] ^{\frac{1}{4}},\label{equation:F5}
\end{align}
where \(T_\text{h}(x,t)=\left[ T_\text{D}(x,t)+T_\text{T}(x,t)\right]/2 \) is the temperature of the hot spot, \(x\) represents the position of the cell in simulations and ``\(\underset{x}{\max}\left\lbrace \cdot\right\rbrace \)'' only perform maximum value operation on \(x\). To match \(u_\parallel\), the \(1/4\) on the exponent here gives the weight the same form as the velocity, derived from \(v\sim\sqrt{T}\). Neutrons are produced through fusion reactions, and the temperature significantly affects fusion reactivity. Thus, directly taking fluid velocity as an estimate of \(u_\parallel\) would erroneously amplify the influence of low-temperature regions, while using temperature-related weights could provide better approximations. It should be mentioned that the influence of fluid flow velocity is significant in the relationship of \(v_\text{iso}\)-\(T_\text{ion}\), and the illustrative determination shown in Fig.\(\ \)\ref{fig:figure2}(c) of the main text did not depict the terms of \(u_\parallel\).

\begin{figure}[htbp]
	\includegraphics[scale=0.26]{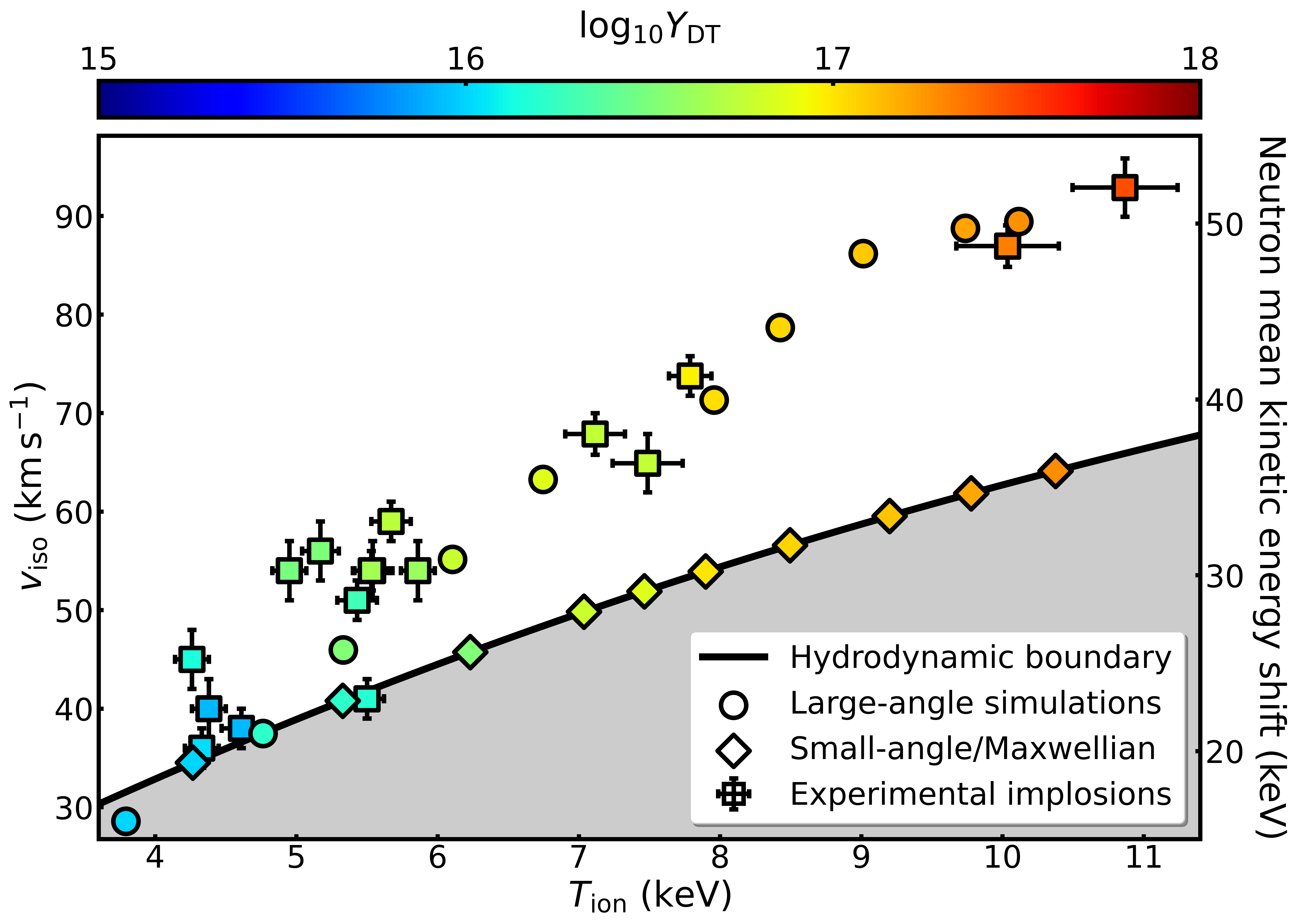}
	\caption{\label{fig:figureS5} The \(v_\text{iso}\)-\(T_\text{ion}\) relationships without translation in simulations after considering large-angle collisions. The sign of deviation between the simulations and the experiments changes when the temperature grows through a temperature around \(9.0\, {\rm keV}\). Moreover, the deviation is more obvious in the low-temperature region where the spherical shock driven implosion process becomes dominant in the fluid velocity \(u_\parallel\).}
\end{figure}

\begin{table*}[htbp] 
\begin{center}
	\begin{threeparttable}
	\caption{\label{tab:tableS1}Initial conditions of the hot spots and values depicted with circle and diamond symbols in Fig.\(\ \)\ref{fig:figure1} and Fig.\(\ \)\ref{fig:figureS5}.}
	\begin{tabular}{m{1.85cm}m{1.5cm}m{1.5cm}m{1.5cm}m{1.5cm}m{1.5cm}m{1.5cm}m{1.5cm}m{1.5cm}m{1.5cm}m{1.1cm}}
		\hline\hline\noalign{\smallskip}\noalign{\smallskip}
		LAPINS & \multirow{2}{*}{3.5k70}  & \multirow{2}{*}{4.0k70}  & \multirow{2}{*}{4.5k70}  & \multirow{2}{*}{5.0k70} & \multirow{2}{*}{5.5k70} & \multirow{2}{*}{6.0k70} & \multirow{2}{*}{7.0k70} & \multirow{2}{*}{8.0k70} & \multirow{2}{*}{9.0k70} & \multirow{2}{*}{10.0k70} \\
		simulations &&&&&&&&&& \\
		\noalign{\smallskip}\hline\noalign{\smallskip}
		\(D_\text{init} \, {\rm (\upmu m)}\) & 70.0000  & 70.0000  & 70.0000  & 70.0000 & 70.0000 & 70.0000 & 70.0000 & 70.0000 & 70.0000 & 70.0000 \\
		\noalign{\smallskip}
		\(T_\text{init} \, {\rm (keV)}\) & 3.5000  & 4.0000 & 4.5000  & 5.0000 & 5.5000 & 6.0000 & 7.0000 & 8.0000 & 9.0000 & 10.0000 \\
		\noalign{\smallskip}
		\(T_\text{ion}^\text{L} \, {\rm (keV)}\) & 3.7935 & 4.7643 & 5.3327 & 6.1063 & 6.7475 & 7.9585 & 8.4266 & 9.0144 & 9.7377 & 10.1158 \\
		\noalign{\smallskip}
		\(T_\text{ion}^\text{S} \, {\rm (keV)}\) & 4.2673 & 5.3287 & 6.2317 & 7.0369 & 7.4648 & 7.8986 & 8.4955 & 9.2010 & 9.7788 & 10.3761 \\
		\noalign{\smallskip}
		\(V_\text{iso}^\text{L} \, {\rm (km/s)}\) & 33.4312 & 41.4228 & 45.0319 & 52.2695 & 60.0469 & 68.1681 & 76.1387 & 85.1654 & 88.8237 & 89.6659 \\
		\noalign{\smallskip}
		\(V_\text{iso}^\text{S} \, {\rm (km/s)}\) & 39.3824 & 44.7436 & 44.8082 & 46.9581 & 48.6905 & 50.7701 & 54.0421 & 58.5338 & 61.9318 & 64.3569 \\
		\noalign{\smallskip}
		\(\tilde{V}_\text{iso}^\text{L} \, {\rm (km/s)}\) & 28.5604 & 37.4673 & 45.9506 & 55.1388 & 63.2506 & 71.3106 & 78.6655 & 86.1699 & 88.7341 & 89.4209 \\
		\noalign{\smallskip}
		\(\tilde{V}_\text{iso}^\text{S} \, {\rm (km/s)}\) & 34.5117 & 40.7881 & 45.7270 & 49.8274 & 51.8942 & 53.9126 & 56.5689 & 59.5383 & 61.8422 & 64.1119 \\
		\noalign{\smallskip}
		\(Y_\text{DT}^\text{L} \, {(10^{16})}\) & 1.00083 & 1.61778 & 3.32093 & 6.83476 & 7.12536 & 10.5523 & 10.9912 & 12.2357 & 16.1071 & 18.7266 \\
		\noalign{\smallskip}
		\(Y_\text{DT}^\text{S} \, {(10^{16})}\) & 1.00103 & 1.62145 & 3.27570 & 6.69955 & 7.26006 & 10.0100 & 11.2799 & 12.5291 & 15.5473 & 19.0462 \\
		\noalign{\smallskip}\hline\hline
	\end{tabular}
\begin{tablenotes}
	\footnotesize
	\item[L/S] Represents the simulation after/before considering large-angle collisions.
	\item[\(\sim\)] Represents data after the translation operation.
\end{tablenotes}
	
	\end{threeparttable}
\end{center}
\end{table*}

The fluid velocity \(u_\parallel\) in simulations mainly came from the deposition of \(\alpha\)-particles around the hot spot. Another source of fluid velocity that affects the \(v_\text{iso}\)-\(T_\text{ion}\) relationship is the residual velocity \cite{mannion2023evidence} of the plasma caused by the spherical shock driven implosion process. The former is more pronounced in the self-sustaining \(\alpha\)-heating process at ion temperatures above approximately \(6.0\, {\rm keV}\), while the latter is the opposite and was not included in the initial simulation condition in this article, which could cause a certain deviation below \(6.0\, {\rm keV}\). Since in large-angle collisions, the \(\alpha\)-particle deposition has minimal impact on the process of the spherical shock driven implosion or the residual velocity of the plasma, it could be considered that this deviation in simulations before and after considering large-angle collisions was similar, which could be eliminated in a translation operation. Moreover, taking into account the limitations of the variance of the small-angle collision algorithm, simulations before considering large-angle collisions should be regarded as Maxwellian simulations.
Thus, in Figure \ref{fig:figureS5}, the points of the \(v_\text{iso}\)-\(T_\text{ion}\) relationship in these Maxwellian simulations were shifted to reach the hydrodynamic boundary as expected, together with the respective points that considered large-angle collisions under the same initial condition. The translation operation only changed the value of \(v_\text{iso}\) without changing the relative relationship between the simulated points before and after considering large-angle collisions. The specific values of \(v_\text{iso}\), \(T_\text{ion}\), \(Y_\text{DT}\), as well as the initial conditions are shown in Table \ref{tab:tableS1}.

The hydrodynamic boundary or the thermal expectation boundary is defined \cite{hartouni2022evidence} by
\begin{equation}\label{equation:F4}
	v_\text{iso}(T_\text{ion})=1.4641\bar{K}+0.37969T_\text{ion},
\end{equation}
where the most probable energy \(\bar{K}\) of the ions involved in fusions \cite{hartouni2022evidence, brysk1973fusion, ballabio1998relativistic, appelbe2014relativistically, munro2016interpreting} can be calculated as 
\begin{eqnarray}\label{equation:F5}
	\bar{K}&=&\frac{\int{{\rm d}K \, K^2\sigma(K)e^{-K/T_{\text{D}}}}}{\int{{\rm d}K \, K\sigma(K)e^{-K/T_{\text{D}}}}} \nonumber \\
	&\to& T_{\text{D}}\left[ \left({T_G}/{T_{\text{D}}}\right) ^{{1}/{3}}+\mathcal{F}(T_{\text{D}})\right] ,
\end{eqnarray}
where \(K\) is the relative kinetic energy of the ions, \(\sigma(K)\) is the fusion reaction cross section,  \(T_G\) is equal to \(295.5\, {\rm keV}\) in deuterium-tritium fusion, and numerical coefficients of the function \(\mathcal{F}(T)\) are from Munro (Table A3 in ref. \cite{munro2016interpreting}), specifically.
\end{document}